\documentclass[article,amsmath,amssymb,twocolumn,superscriptaddress]{revtex4}
\usepackage[subpreambles=true]{standalone}
\usepackage{import}

\usepackage{blindtext}
\usepackage{chngcntr}
\usepackage{grffile}
\usepackage{MnSymbol}
\usepackage{textcomp}
\usepackage[dvipsnames]{xcolor}
\usepackage{xcolor}	
\usepackage{amsmath,color}

\usepackage[colorlinks,
            linkcolor=blue,
            anchorcolor=blue,
            citecolor=blue
            ]{hyperref}

\begin{document}

\title{Tail-induced equilibration in long-range interacting quantum lattices}
\author{Wei-Han Li}\email{weihan@outlook.com}
\affiliation{Semmelweistr. 2, 01159 Dresden, Germany}
\begin{abstract}
We examine the relation between inter-particle interactions and real-time equilibration in one-dimensional lattice systems with hard-core constraints. Focusing on the roles of interactions, our results demonstrate that in the presence of interaction tails, any power-law exponent (including the limit ones) can encode the random particle configurations to the Hamiltonian, leaving the latter characterized by random matrices. Through an experimental-accessible setup using dipolar-interacting particles in optical lattices, the quenched relaxations are demonstrated resulting in equilibrium, and the relation between eigenstate thermalization is confirmed. Our study directly unveiled the role of inter-particle interactions in quantum many-body dynamics, offering a new scheme to address equilibration in closed quantum many-body problem based on the manifesting of random particle configurations in the model Hamiltonian. 
\end{abstract}
\maketitle

\section{Introduction}
\label{intro}

A quantum system with inter-particle interactions is believed to thermalize according to the eigenstate thermalization hypothesis (ETH) \cite{Rigol2008, Deutsch1991, Srednicki1994, DAlessio2016}, while the most crucial exception occurs in various one-dimensional (1D) systems due to the integrability \cite{int}. Investigations concerning integrability and its breaking are mostly based on systems belonging to "short-range" lattice models with the interactions or couplings between particles being only up to next-to-nearest neighbors. Such models include the Bose-Hubbard model \cite{BH} and its extension \cite{EBH}, spin models of Ising \cite{Ising} or XXZ \cite{XXZ} types, $t$-$V$ models for hare-core bosons or spinless fermions \cite{BF_Rigols, BF_others}, and photon-particle mixed models \cite{mixed_model}.

Recent experiments are focusing on "long-range" interacting lattice systems \cite{Defenu2023}, where strong inter-site interactions can be engineered in optical lattices. In particular, experiments using magnetic atoms \cite{Paz2013, Patscheider2020, Baier2016, Su2023} and Rydberg gases \cite{Scholl2021, Sanchez2021} have already realized long-range spin models \cite{Paz2013, Patscheider2020, Scholl2021} and extended Hubbard models (EHMs) \cite{Baier2016, Su2023, Sanchez2021}. Intriguing nonequilibrium dynamics have been predicted in EHMs \cite{Barbiero2015, WHL1, WHL2, WHL3, Korbmacher2023}, provide that the inter-site interaction is strong enough such that the Hilbert space is shattered or fragmented. In contrast, such systems are expected to equilibrate through weaker interaction strength that can be realized much more easier in the above experiments. But as far as we know, this expectation has not been justified systematically. 

Here we consider the EHM consisting of hard-core bosons, or spinless fermions, with dipole-dipole (DD) interactions. The tail effect beyond the nearest neighbor (NN) is confirmed to result in equilibration in far-from-equilibrium dynamics. Moreover, although there is no external random field and the lattices are homogeneous, we demonstrate that the Hamiltonian is described by random matrix theory (RMT), and illustrate the relationship with ETH using the eigenvector complexity. We show that the randomness originates from the arbitrary particle configurations, which does not depend on the specific value of the interaction exponents. This implies that our results can be expanded to general long-range interacting systems. 

\section{The model}
\label{model}

We consider the quantum lattice gases described by
\begin{equation}\label{H}
H_{DD} = H_{NN} + H_t,
\end{equation}
where $H_{NN} = \sum_{j} \left[ -t\left( \hat{a}^{\dagger}_{j+1}\hat{a}_{j} + \textrm{H.c.} \right) + V\hat{n}_{j+1} \hat{n}_j \right]$ is the nearest-neighbor model, and $H_t = V \sum_{i+1<j} \frac{ \hat{n}_{j}\hat{n}_{i} } {(j-i)^3}$ is the part of tail. The operator $\hat{a}^{\dagger}_j$ ($\hat{a}_j$) creates (annihilates) a particle at lattice site $j$ and $\hat{n}_j = \hat{a}^{\dagger}_j\hat{a}_j$ is the on-site occupation number with the hard-core constraint $(\hat{a}^{\dagger}_j)^2=0$. 
To align with experiments, we consider open boundary conditions (OBC) for $N$ particles moving in $L$ sites with  $N/L=\frac{1}{2}$. 
The quantum lattice gas model can be mapped to the spin model through the Holstein–Primakoff transformation, thereby $H_{NN}$ is isomorphic to the XXZ model (the spin-$\frac{1}{2}$ Heisenberg chain) \cite{note_mapping1}. Our strategy is to unveil the effects of the tail by comparing the far-from-equilibrium dynamics and the eigenstates of $H_{DD}$ to those of $H_{NN}$. We will let $V/t = 2$, this value is easily accessible in the present experiments, and it leads to a well-known spectrum level statistics after mapping to the XXZ model \cite{Poilblanc1993}.


\begin{figure}[t]
\includegraphics[width=\columnwidth]{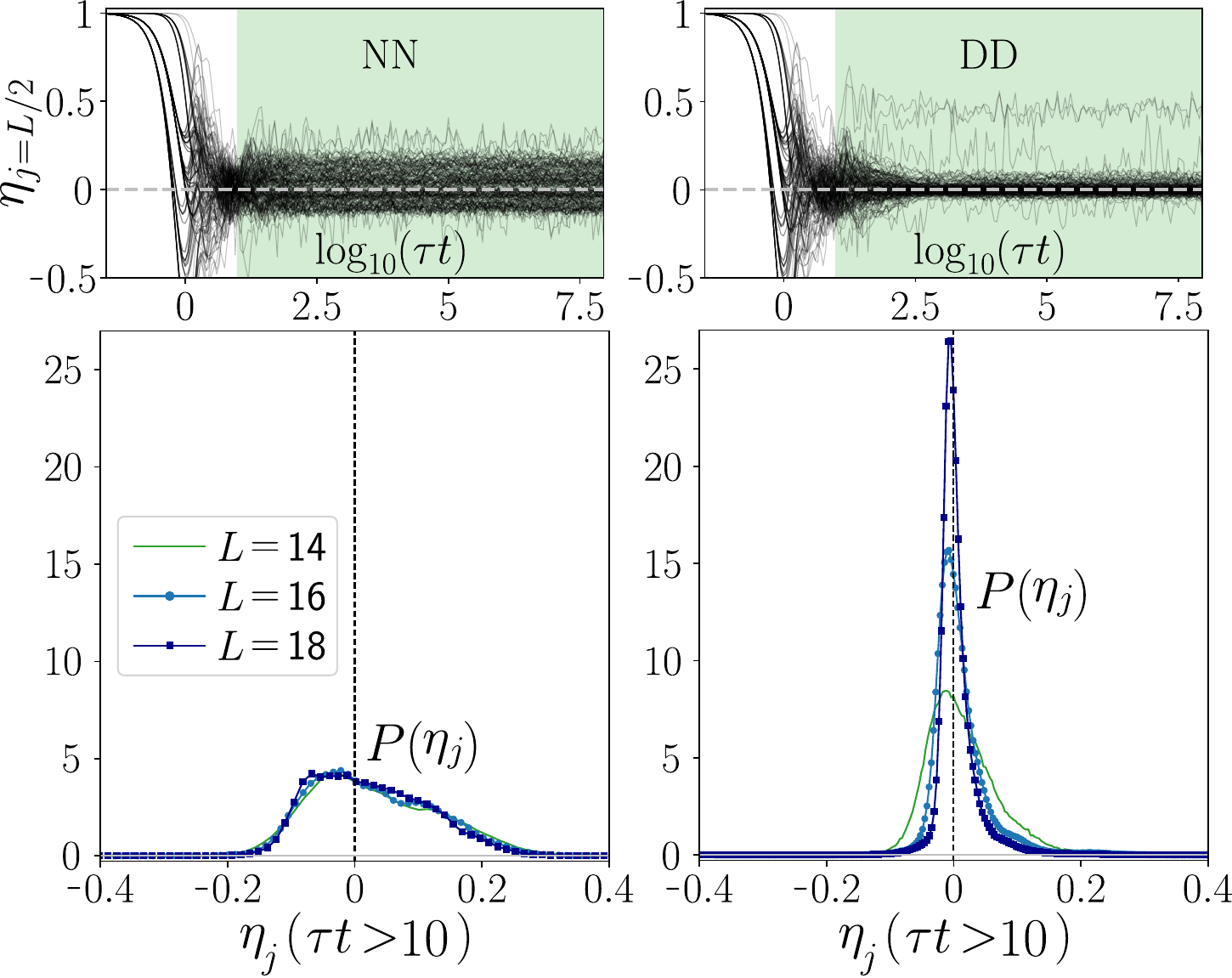} \centering
	\caption{Real-time dynamics and statistics of equilibration. Upper panels show the local equilibration parameter $\eta_{j}(\tau)$ for initial Fock states $|f_j\rangle$ with $L=16$ and $j=L/2$, only $1$ out of every $40$ evolution are plotted, significant difference reveals between $H_{NN}$ and $H_{DD}$ at $\tau t \gtrsim 10$. Lower panels show the long-time probability density distribution $P(\eta_j)$ over all initial states for various $L$. Though not shown here, $P(\eta_j)$ for individual initial state behaves similarly to the overall $P(\eta_j)$.} 
	\label{fig_1}
\end{figure}

\section{Tail-induced equilibration}
\label{equilibration}
We examine the dynamical equilibration of Fock states $|f_j\rangle$ with the $j^{\textrm{th}}$ site initially occupied, these initial conditions can be prepared using atomic microscopes \cite{Bakr2009, Kuhr2016, Gross2021}, 
enabling site-resolved single-atom detection. Focusing at the center of the lattices, i.e., $j=L/2$, we compute 
\begin{equation}\label{eta}
\eta_{j=L/2}(\tau) = \frac{\langle f_j(\tau)|\hat{n}_{j}| f_j(\tau) \rangle-N/L}  {1-N/L}.
\end{equation} 
The parameter $\eta=1$ indicates that the initially observed particle is localized at $j$, while $\eta=0$ implies that it has dissipated into the background. The time evolution is implemented using eigenstates of Eq. (\ref{H}) obtained by exact diagonalization (ED). 
Figure \ref{fig_1} shows $\eta_{j=L/2}(\tau)$ in a long timescale, the results of $H_{DD}$ compared with that of $H_{NN}$ show that the dipolar tail leads to evidently stronger equilibration at $\tau t \gtrsim 10$ (upper panels). We applied the probability density $P(\eta_j)$ for $\eta_j$ at a long timescale (lower panels) to quantify the tail-induced equilibration. For $H_{DD}$, the distribution of $P(\eta_j)$ narrows and peaks at $\eta_j = 0$ as $L$ increases. In contrast, $P(\eta_j)$ is significantly flatter for $H_{NN}$, although a slightly concentrating is also observed when $L$ increases. This indicates that the tail diminishes the border effects that prevent equilibration in a small quantum lattice gas.

\section{Random Matrix theory}
\label{RMT}
RMT \cite{Mehta2004} provides, from the viewpoint of eigenstates, a good understanding of the physics underlying our dynamics. If the allowed energy of a system behaves as an independent random variable, the statistics of energy levels give the Poisson distribution: $P(s) = e^{-s}$. However, if the energy belongs to eigenvalues of a random matrix, that is, if the Hamiltonian is described by a matrix with random elements, then the level statistics is believed to be the Wigner-Dyson (W-D) distribution \cite{Guhr1998, Mehta2004, Kravtsov2012}. For a Hamiltonian with time-reversal symmetry, the matrix elements are real random variables captured by the Gaussian orthogonal ensemble (GOE), leaving the W-D distribution as $P(s) = s\, e^{-s^2}$. 

\subsection*{Level statistics}
\label{level}
\begin{figure}[b]
\includegraphics[width=\columnwidth]{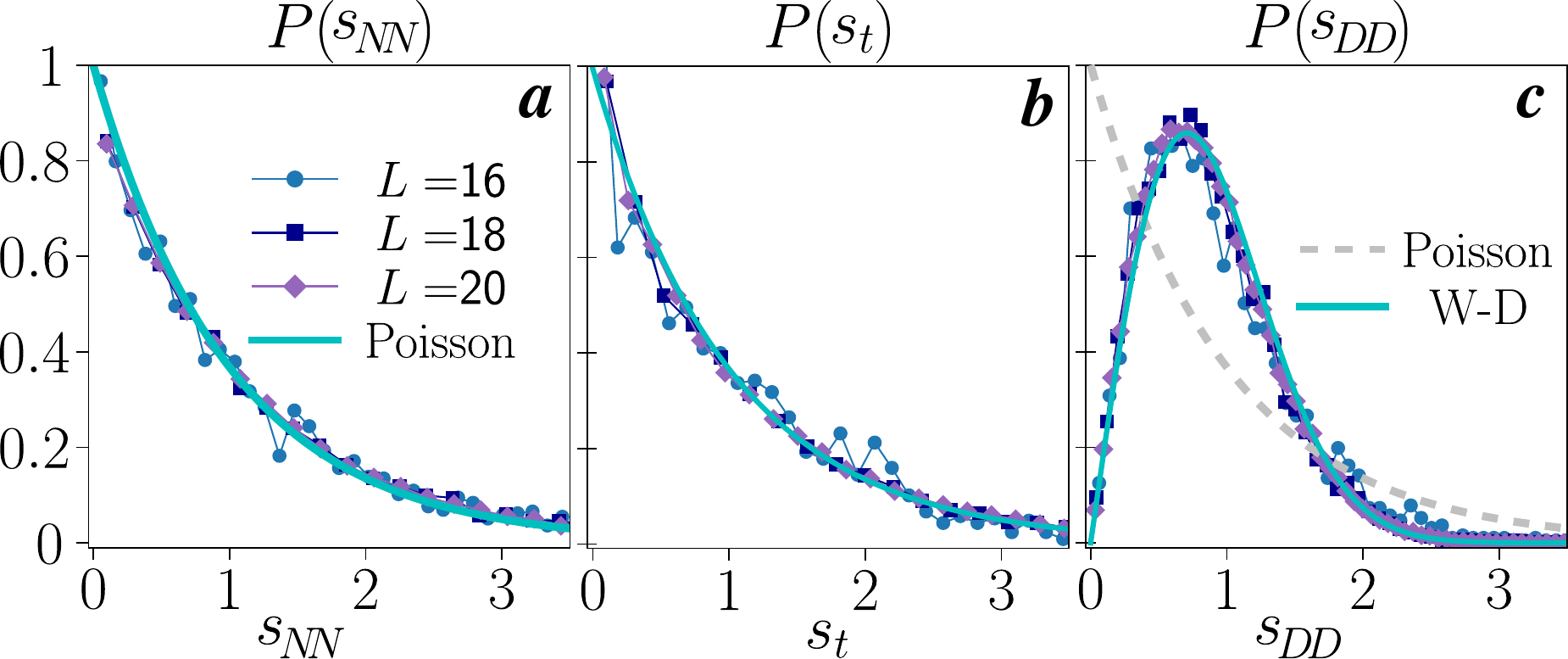} \centering
	\caption{Level statistics for various $L$. The probability density $P(s)$ of the re-scaled level spacing $s$ (see the main text) are given for  (\textit{\textbf{a}}) the NN model $H_{NN}$,  (\textit{\textbf{b}}) the tail energy $E_t$, and (\textit{\textbf{c}}) the full system $H_{DD}$. The solid curves fit the data, and the dashed curve is added for comparison.}
	\label{fig_2}
\end{figure}

Considering the orthogonal transformation: $O^{T} H_{DD} O = E_{NN} + O^{T} H_t O$, where $O$ and $E_{NN}$ are matrix of normalized eigenvectors and eigenvalues of $H_{NN}$ respectively. Recalling that $H_{NN}$ is integrable as it can be mapped to the XXZ model, whose eigenstates can be solved using Bethe ansatz methods \cite{int}. Previous studies have shown that $E_{NN}$ behave as independent random variables, the probability density of the level spacing between consecutive eigenvalues reveals a Poisson distribution \cite{Poilblanc1993, Berry1977}. Our simulations shown in Fig. \ref{fig_2} (\textit{\textbf{a}}) confirm this result. 

The numerical method is detailed as follows. The ED is implemented with parity and particle-hole symmetry, and the data is presented in the sub-space of odd parity and even particle-hole symmetry \cite{footnote1}. 
We obtained the energy spectrum in the sub-space for the spin model corresponding to either $H_{NN}$ or $H_{DD}$. We divided the spectrum into small intervals, and within an interval that contains, say, $E_1 \leq E_2 \cdots \leq E_X$, we calculate the averaged spacing $s_n = S_n/\overline{S} $ with $S_n = E_{n+1}-E_n$ the consecutive level spacing and $\overline{S}$ the averaged spacing within the interval. The probability density $P(s)$ is obtained with $s$ in the middle of the spectrum, the data that are discarded close to the border of the spectrum is $1-5\%$. 
Though not shown here, the statistics in different sub-spaces are also checked to give the same results.

The term $H_t$ encodes random particle configurations in the system through the tail of dipole-dipole interactions, after the orthogonal transformation, $O^{T} H_t O$ contributes as random off-diagonal terms of the Hamiltonian matrix. To examine this randomness, we apply level statistics to the tail energy $E_t = \langle f|H_t|f \rangle$ with $|f \rangle$ denoting all the Fock bases. Moreover, to focus on the randomness embedded in particle configurations, the configurations being identical under reflection and shifting should be avoided. 
Fig. \ref{fig_2} (\textit{\textbf{b}}) demonstrates this expectation by the Poissonian distribution $P(s_t) = e^{-s_t}$. 
We hence argue that $O^T H_{DD} O = E_{NN} + O^T H_t O$ behaves as a random matrix, and according to RMT, $P(s_{DD}) = s_{DD}\, e^{-s_{DD}^2}$. Fig. \ref{fig_2} (\textit{\textbf{c}}) justifies this expectation. Moreover, it has been shown that if the Hamiltonian is a random matrix, then the RMT production of observables leads to thermalization in the sense of ETH \cite{DAlessio2016}. 

\subsection*{Randomness from the particle configurations}
\begin{figure}[b]
\includegraphics[width=\columnwidth]{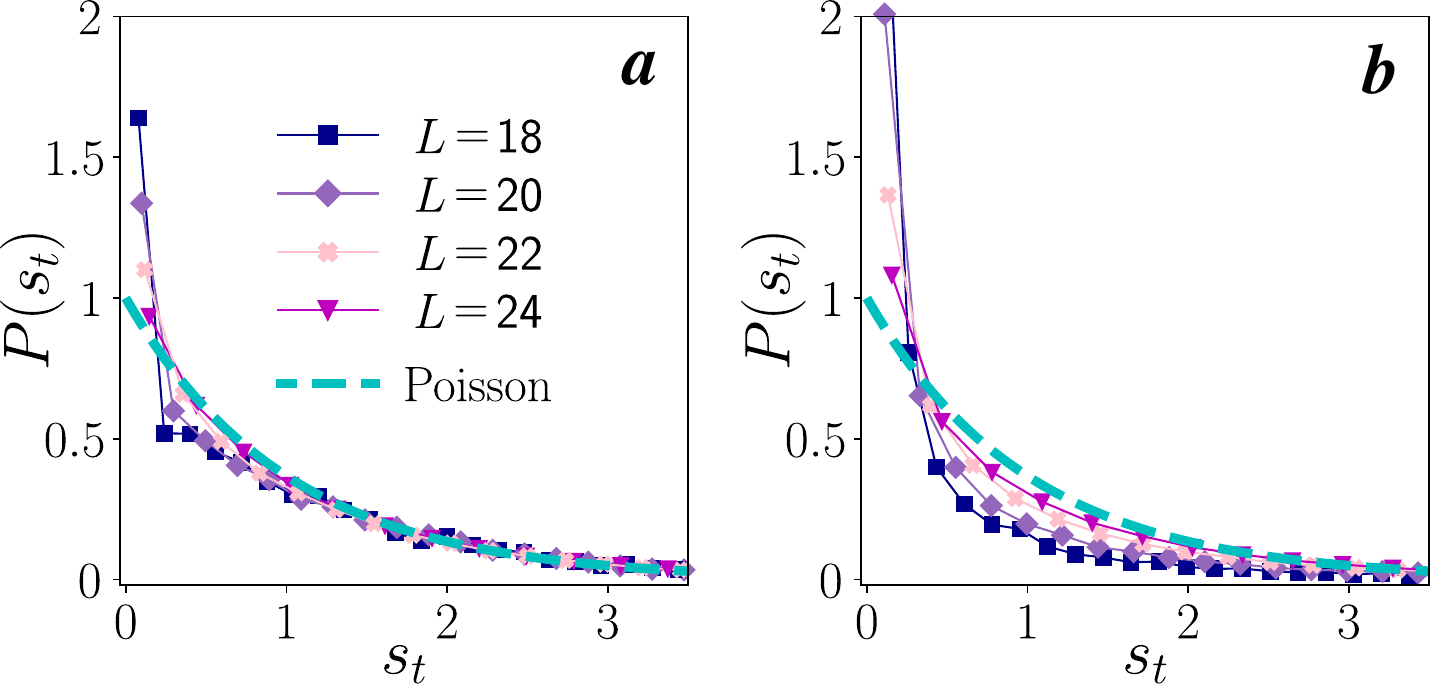} \centering
	\caption{Randomness originating from particle configurations. As represented by the level statistics of the tail energy $E_t(\alpha)$, the randomness is confirmed when $P(s_t)$ agrees with the Poisson distribution. (\textit{\textbf{a}}) and (\textit{\textbf{b}}) show for the $\alpha=10^{-4}$ and $\alpha=9$ respectively, that $P(s_T)$ turns Poissonian as $L$ increases.}
	\label{fig_4}
\end{figure}

The randomness of our model originates from the arbitrary configurations of the particles, which manifests itself through the tail energy $E_t = \langle f|H_t|f \rangle$. Here we examine how this manifesting depends on the tail's length exponent. Considering $H_t(\alpha) = \sum_{i+1<j} \hat{n}_i\hat{n}_j/(j-i)^{\alpha}$, if the tail energies act as independent random variables and the randomness is well represented when $\alpha$ varies, then the Poissonian $P(s_t)$ is expected even at $\alpha \to 0$ and $\alpha \to \infty$, even though the former approaches the non-interacting case and the latter approaches the NN model. Otherwise, the Poisson reduces to a Dirac delta function at the two limits. Fig. \ref{fig_4} shows the $P(s_t)$ with numerically extreme $\alpha$. For a small system, a delta-like distribution seems to appear, but the distribution quickly returns to the Poisson distribution as $L$ increases. This indicates that the complexity of particle configuration can sufficiently overwhelm the coincident degeneracy. Exponents with moderate values result in a faster approach to the Poisson distribution. 

As discussed above, a random $H_t(\alpha)$ suggests that the Hamiltonian represented by $E_{NN} + O^T H_t(\alpha) O$ is described by RMT, which is followed by ETH \cite{DAlessio2016}. Hence, the equilibrium of $\hat{n}_{j=L/2}(\tau)$ is expected on a long time scale. 
Note that here the interaction strength $V$ is irrelevant as the statistics are carried out in the level spacing re-scaled by the local average: $s_T^{n}=S_T^{n}/\overline{S_T}$. Furthermore, the off-diagonal term $O^T H_t(\alpha) O$ guaranteed level repulsion, i.e., $P(s)\to 0$ for $s\to 0$, even if the tail is a perturbation with extreme $\alpha$, but further analysis is needed to justify if $P(s)$ approaches the W-D distribution. 

\section{Eigenvector complexity and ETH}
\label{eigenvector}
RTM is also applied to analyze eigenvectors \cite{Izrailev1990, Zelevinsky1996, Kota2001} and leads to important features of ETH via the complexity of eigenvectors in the Hilbert space \cite{BF_Rigols, Zelevinsky1996}. 
We apply Shannon entropy to measure the complexity, representing the eigenvectors $|E\rangle$ in the Fock bases $|f\rangle$, we calculate
\begin{equation}
    \label{Shannon}
    S(E) = -\sum_{f} |\langle f|E \rangle|^2 \ln{ |\langle f|E \rangle|^2 }.
\end{equation}
ETH states that the thermalization of an observable occurs in an isolated quantum system when eigenstates being close in energy are also thermal, such that they give close expectation values of that observable. According to Percival's conjecture \cite{Percival1973}, eigenstates adjacent in energy are very similar, one can expect that they are severely mixed superpositions of the bases \cite{Zelevinsky1996} and their Shannon entropy essentially approaches the same, hence a smooth function of $S(E)$ is expected along with ETH.
\begin{figure}[b]
\includegraphics[width=\columnwidth]{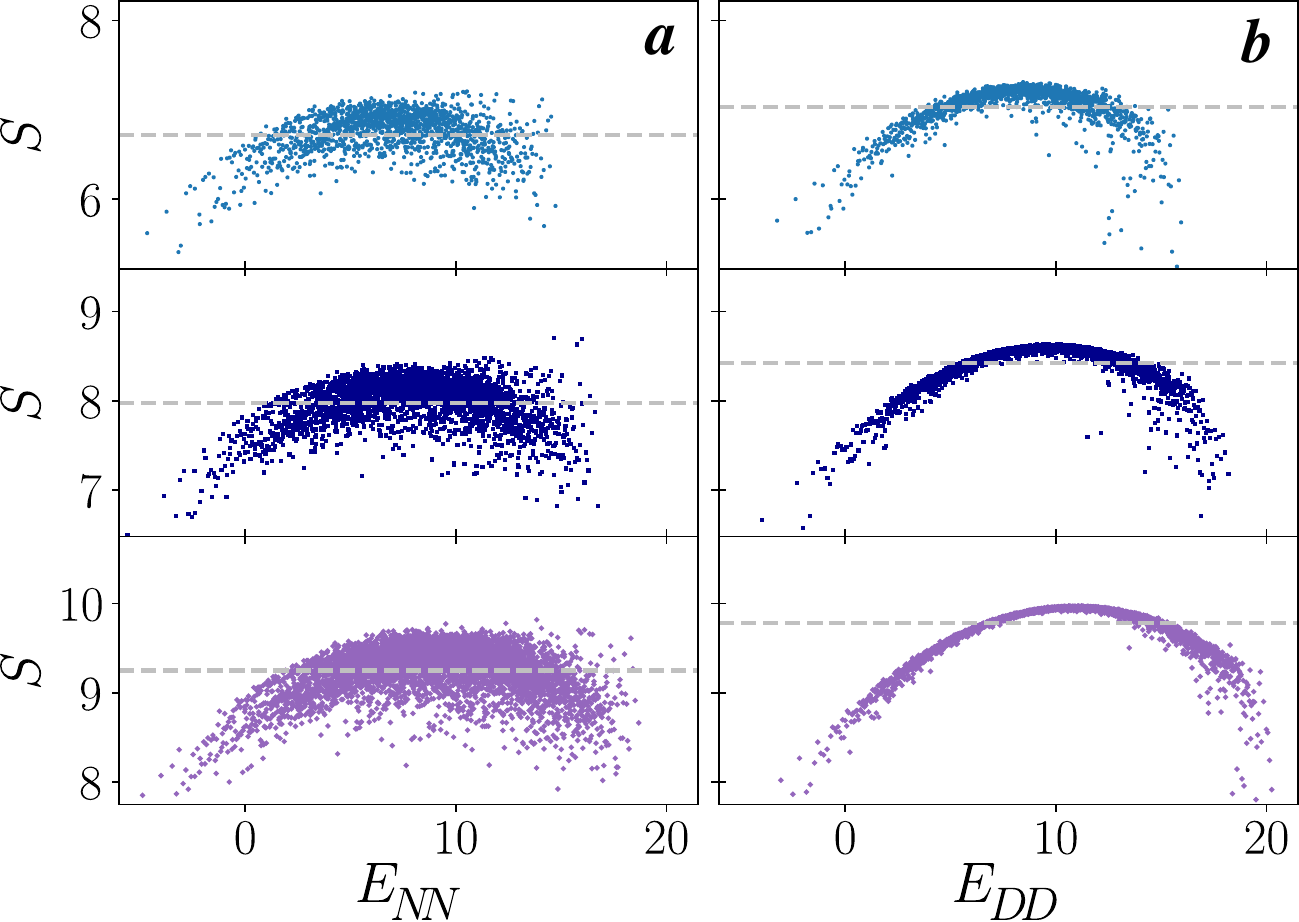} \centering
	\caption{ Eigenstate complexity represented by the Shannon entropy $S$ for eigenvectors of $H_{NN}$  (\textit{\textbf{a}}) and $H_{DD}$ (\textit{\textbf{b}}) in the Fock state bases. From top to bottom, $L=16, 18, 20$ and only $1$ out of $2$, $4$, $8$ points are shown. The dashed lines denote the averaged values of $S$.}
	\label{fig_3}
\end{figure}

As shown in Fig. \ref{fig_3} (\textit{\textbf{a}}), the strong fluctuations of $S(E_{NN})$ indicate a high degree of difference of complexity within a narrow energy interval. The lack of similarity between the eigenvectors implies poor compliance with ETH, this explains the absence of ultimate equilibration in Fig. \ref{fig_1} (left panels). In contrast, fluctuations of $S(E_{DD})$ reduce in the middle of the spectrum, it approaches a smooth function of $E_{DD}$ as $L$ increases. This implies that eigenvectors close in energy are the combinations of bases coupled by the tail with similar magnitude, increasing entropy in the middle of the spectrum further showing that the involved bases increase numerously with the density of states. These signals of ETH explain the significant equilibration shown in Fig. \ref{fig_1} (right panels). In the context of ETH, the matrix elements of the observable in the eigenstate bases are also characterized by smooth functions of energy, this has been studied for $\hat{n}_{j=L/2}$ in polar lattice gases similar to Eq. (\ref{H}) and fluctuation-dissipation theorem is confirmed \cite{Khatami2013}. 
Fluctuation increases at the border of the spectrum, where the density of states approaches zero—implying that for dynamics close to the ground states and the highest excited states, ETH fails and equilibration is absent. This is seen by the minor cases of $\eta_j \neq 0$ in the long timescale in Fig. \ref{fig_1} (right panels).

The values of Shannon entropy depend on the bases chosen for the representation, here we take the Fock bases for the study of equilibration in real space as shown in Fig. \ref{fig_1}. But as eigenstates satisfying ETH have no good quantum numbers in addition to the energies, the severely mixed eigenvectors in the chosen bases remain mixed in other bases. The complexity quantity, $S(E)$, is thus expected to remain smooth in the middle of the spectrum, though the values change with representations.

\section{Conclusions}\label{discussion}
We demonstrate that the random matrix theory can describe the Hamiltonian of a clean lattice system consists polar gases aligning with the current experiments, furthermore, the eigenvectors are captured by the eigenstate thermalization hypothesis and far-from-equilibrium particle configurations equilibrate as a consequence. Although previous studies have shown that the next-to-nearest neighbor coupling, including hopping and interaction, can give rise to the random matrix theory and eigenstate thermalization \cite{BF_Rigols, Poilblanc1993}. Here we demonstrated the effects originate exclusively from the interactions (or, the Ising terms when mapping into spin models). The physical origin for the randomness satisfying the random matrix theory is the arbitrary particle configurations, which do not depend on the specific value of the interaction exponent. The manifesting of random particle configurations in the model Hamiltonian provides an intuitive perspective to the issue of quantum dynamical equilibration, offering a new scheme for studying equilibration in close quantum many-body systems.



\begin{thebibliography}{99}
\bibitem{note_mapping1} The Holstein–Primakoff transformation mapping is given as $\hat{a}_j \mapsto \frac{1}{2} (\sigma_j^{x} - i\sigma_j^{y})$ and $\hat{n}_j \mapsto \frac{1}{2}(\sigma_j^{z} +1) $, where $\sigma^{x,y,z}$ are Pauli matrices. It results in $H_{NN} \mapsto H_{XXY}+c$ for periodic boundary conditions, but $H_{NN} \mapsto H_{XXY}+\frac{1}{2}(\sigma_1 + \sigma_L)+c$  for open boundary conditions.
\bibitem{footnote1} Note that the original EHM, Eq. (\ref{H}), has no particle-hole symmetry due to the OBC, we have mapped the EHM into the spin model and ignored the edge states \cite{note_mapping1}. The resultant spin model reveals spin-flip symmetry, of which we take advantage. Level statistics of the EHM and the spin model are checked to be the same with available system sizes.
\end{thebibliography}
\end{document}